\documentclass[usenatbib,fleqn,useAMS]{mnras}
\usepackage{newtxtext,newtxmath}
\usepackage[T1]{fontenc}
\usepackage{ae,aecompl}
\usepackage[pdftex]{graphicx}
\usepackage{epstopdf}

\usepackage{amsmath}
\usepackage{amssymb}	
\usepackage{color}

\usepackage{graphicx}

\usepackage{natbib}
\usepackage{aas_macros}
\usepackage[section] {placeins}
\usepackage{subfigure}
\usepackage{xspace}
\usepackage{float}
\usepackage{color}
\usepackage[dvipsnames]{xcolor}

\def\msun{{\rm\,M_\odot}}

\def\msun{{\rm\,M_\odot}}

\newcommand{\kms}{\, {\rm km\, s}^{-1}}

\newcommand{\be}{\begin{equation}}
\newcommand{\ee}{\end{equation}}

\def\h2{${\rm\,H_2}$}

\usepackage{color}

\newcommand{\vmax}{v_{\rm max}}
\newcommand{\maxm}{\texttt{maxm}}
\newcommand{\hargism}{\texttt{h14m}}
\newcommand{\hargisr}{\texttt{h14r}}
\newcommand{\hargisi}{\texttt{h14i}}
\newcommand{\zpeak}{z_{\rm peak}}
\newcommand{\vpeak}{v_{\rm peak}}
\newcommand{\Mpeak}{M_{\rm peak}}

\title[Selecting UFD hosts in N-body sims at high z]{Selecting ultra-faint dwarf candidate progenitors in cosmological N-body simulations at high redshifts}
\author[Safarzadeh, Ji et al.]{
	\parbox[t]{\textwidth}{
	Mohammadtaher Safarzadeh$^{1}$\thanks{E-mail: mts@asu.edu}, Alexander P. Ji$^{2,\dag}$, Gregory A. Dooley$^{3}$ , Anna Frebel$^{3}$, Evan Scannapieco$^{1}$, 
	Facundo A. G\'omez$^{4,5}$, Brian W. O'Shea$^{6,7,8,9}$} \vspace*{12pt}\\
	$^{1}$School of Earth and Space Exploration, Arizona State University, Tempe, AZ 85287-1404, USA\\
	$^{2}$The Observatories of the Carnegie Institution of Washington, 813 Santa Barbara St., Pasadena, CA 91101, USA\\
	$^{\dag}${Hubble Fellow}\\
	$^{3}$ Department of Physics, Kavli Institute for Astrophysics and Space Research, Massachusetts Institute of Technology,\\
	77 Massachusetts Avenue, Cambridge, MA 02139, USA\\
	 $^{4}$Instituto de Investigaci{\'o}n Multidisciplinar en Ciencia y
	Tecnolog{\'i}a, Universidad de La Serena, Ra{\'u}l Bitr{\'a}n 1305, La Serena, Chile\\
	$^{5}$Departamento de F{\'i}sica y Astronom{\'i}a, Universidad de La
	Serena, Av. Juan Cisternas 1200 N, La Serena, Chile\\
	$^{6}${Department of Computational Mathematics, Science and Engineering, Michigan State University, MI, 48823, USA}\\
	$^{7}${Department of Physics and Astronomy, Michigan State University, MI, 48823, USA}\\
	$^{8}${National Superconducting Cyclotron Laboratory, Michigan State University, MI, 48823, USA}\\
	$^{9}${Joint Institute for Nuclear Astrophysics - Center for the Evolution of the Elements, USA}\\
}

\usepackage[all]{hypcap}

\pubyear{2017}

\begin{document}
\label{firstpage}
\pagerange{\pageref{firstpage}--\pageref{lastpage}}
\maketitle

\begin{abstract}
The smallest satellites of the Milky Way  ceased forming stars during the epoch of reionization and thus provide archaeological access to galaxy formation at $z>6$.
Numerical studies of these ultra-faint dwarf galaxies (UFDs) require expensive cosmological simulations with high mass resolution that are carried out down to $z=0$.  However, if we are able to statistically identify UFD host progenitors at high redshifts \emph{with relatively high probabilities}, we can avoid this high computational cost. To find such candidates, we analyze the merger trees of Milky Way type halos from the high-resolution ${\it Caterpillar}$ suite of dark matter only simulations. Satellite UFD hosts at $z=0 $ are identified based on four different abundance matching (AM) techniques. All the halos at high redshifts are traced forward in time in order to compute the probability of surviving as satellite UFDs today.  Our results show that selecting potential UFD progenitors based solely on their mass at z=12 (8) results in a 10\% (20\%) chance of obtaining a surviving UFD at $z=0$ in three of the AM techniques we adopted. We find that the progenitors of surviving satellite UFDs have lower virial ratios ($\eta$), and are preferentially located at large distances from the main MW progenitor, while they show no correlation with concentration parameter.  Halos with favorable locations and virial ratios are $\approx 3$ times more likely to survive as satellite UFD candidates at $z=0.$

\end{abstract}

 \begin{keywords}
Galaxies: Dwarf; Galaxies: Local Group ; Galaxies: statistics
\end{keywords}

\section{Introduction}
The faintest satellite galaxies of the Milky Way (MW) preserve an archaeological snapshot of galaxy formation and chemical enrichment in the early universe.
These ultra-faint dwarf galaxies (UFDs) have luminosities of only $L\approx10^{2.5-5}L_{\odot}$ \citep[e.g.][]{McConnachie:2012fh}, but their stars display large velocity dispersions implying they reside in dark matter halos of mass $\sim10^7-10^9\msun$ \citep{Simon:2007ee,Strigari:2008in,Simon:2011dm}.
The stellar populations of UFDs are all very old \citep[$>12\,$Gyr][]{Brown:2014jn,Weisz:2014cp} and metal-poor \citep[$\mbox{[Fe/H]} <-2$, e.g.][]{Kirby:2008fs,Frebel:2012ja, Vargas:2013ei}, implying that UFDs formed nearly all of their stars prior to reionization \citep[e.g.][]{Bullock:2000bn,Bovill:2009hg,Bovill:2011bk}. Observations of the population of UFDs are not complete \citep{Tollerud:2008eq,Hargis14}, and are still being discovered in the Milky Way \citep{DrlicaWagner:2015gb,Koposov:2015cw} and other nearby galaxies \citep{Lee:2017hu}.

The star formation history of UFDs and their host halo characteristics and accretion history have been studied through various techniques \citep{Gnedin:2006gl,Salvadori:2009hs,Bovill:2011bk,GK14a,Wheeler:2015fm,BlandHawthorn:2015ke,GarrisonKimmel:2017cv,Sawala:2017jp,Jeon:2017wo}. However, direct cosmological hydrodynamic simulations to study UFDs remain challenging. 
Zoom simulation boxes with comoving sizes of $\approx10~$Mpc/h on a side \citep{BoylanKolchin:2016ex}, embedded in larger cosmological box sizes, would be required in order to self-consistently simulate the MW in its environment and also to capture the reionization's impact on the star formation history of galaxies due to the patchiness of the process. With such box sizes, a full radiative-hydro simulation with a dark matter particle mass of $\leq 10^4\msun$ needed to resolve UFD host halos requires approximately 10 Million core-hours  to simulate down to redshifts $z\approx8-10$ \citep{Xu:2016gl}. Therefore, it is currently impossible to directly simulate such boxes down to $z=0$ to identify the survived halos with star formation histories resembling UFDs around the Milky Way (though isolated UFDs can still be studied; e.g. \citealt{Wheeler15,Jeon:2017wo}). In this situation, it is useful to have a method to predict the fate of the halos given what we know about
their properties at high redshifts.

The fate of halos given their mass has been studied in the literature in the context of classical dSph galaxies when the satellite to host halo mass ratio is  $\rm m_{sat}/M_{host} > 10^{-3}$. 
\citet{Sales:2007kw} studied the characteristics of the halos that merge into the main host and those that survive in a MW type halo and find that mass and eccentricity of the orbit of the satellite determines the fate of a satellite through dynamical friction \citep{BoylanKolchin:2008ex}. They find that satellites that merge into the host tend to have more eccentric orbits compared to those that survive. Moreover, the merging satellites tend to be accreted into the main host at earlier times, meaning they spend more time within the virial radius of the host and experience dynamical friction on a longer timespan. Likewise, \citet{Klimentowski:2010fu} find the survival probability depends on the number of times a satellites experiences a pericenter passage. 

In this paper, we focus on statistics of the surviving halos given their mass and other characteristics that tie them to the UFDs.
We analyze the merger trees of MW analogs constructed based on N-body simulations to study the characteristics of the halos at high redshifts ($z>4$) that survive as 
UFD host halos today, corresponding to a satellite-to-halo mass ratio of $\sim 10^{-5}$.
Such study would require a particle mass of $\approx 10^4 \msun$ to at least resolve halos of mass $M\approx10^7\msun$ with 1000 dark matter particles. 
The {\it Caterpillar} simulation suite \citep{Griffen16b,Griffen16a} is a set of zoom-in N-body simulations on 32 MW-type halos that provides us the tools for such study.
We identify UFDs at $z=0$ based on various abundance matching (AM) techniques that we describe in the methods section. 
We first focus on the halo mass as the primary informant of survivability of the halos as UFD hosts today. We study what fraction of the halos at a given mass and redshift bin 
(a) survive as UFD hosts, (b) merge into the main host, (c) merge into a classical dSph galaxy, and (d) merge into a UFD candidate halo, i.e., not the main progenitor of the UFD host halo.
We then consider other dynamical parameters such as distance from the main progenitor, virial ratio ($\eta\equiv2T/|U|$) where $T$ and $U$ are the total kinetic and potential of a halo, and concentration parameter of the halos to explore if these parameters drive the survivability of the halos as UFD hosts today.

In \S2 we describe our method in more detail. In \S3 we present our results, and in \S4 we discuss our results and give conclusions. 

\section{method}
\subsection{Simulations}

We use 32 $N$-body zoom-in simulations of Milky Way analogs from the {\it Caterpillar} simulation suite \citep{Griffen16b,Griffen16a}.
These host halos were selected from a 100 $h^{-1}$ Mpc$^3$ box to have 
masses and formation histories consistent with that of the Milky Way ($0.7 \times 10^{12}\msun \leq M_{\rm vir} \leq 3 \times 10^{12} \msun$) and were re-simulated to $z=0$  with particle mass $m_p={\approx}3\times 10^4\msun$, and comoving softening length of 113 pc.
Initial conditions were generated from {\sc music} \citep{Hahn11} for a Planck 2013 cosmology \citep[$\Omega_m=0.32,\ \Omega_\Lambda=0.68,\ n_s=0.96,\ \sigma_8=0.83,\ h=0.6711$][]{Planck14}. The simulations were run with {\sc gadget-3} and {\sc gadget-4} \citep{Springel05}.
The {\sc rockstar} halo finder (\citealt{Behroozi13a} with iterative unbinding implemented as described in \citealt{Griffen16a}) was used and the associated merger trees was constructed with {\sc consistent-trees} \citep{Behroozi13b}. 
In this paper, our analysis is conducted entirely on these merger trees.
For more details, see \citet{Griffen16a}.

Virial halo masses are defined by the \citet{Bryan98} definition.
We refer to the main Milky Way halo ($M_{\rm vir} \approx 10^{12} \msun$) in each simulation as the ``host halo.''
Most of the 32 zoom-in simulations have volumes uncontaminated by low-resolution particles out to 1 Mpc at $z=0$, but throughout this paper we only consider halos that end up within the Milky Way's virial radius at $z=0$.
All ``peak'' quantities are defined as occurring at the snapshot where a halo's main branch reaches its maximum mass; i.e. $\Mpeak$ is the maximum mass of a halo over its main branch, and $\vpeak$ is $\vmax$ at that time.
The virial ratio is defined as $\eta = 2T/|U|$, where $T$ and $|U|$ are kinetic and gravitational energies of the halos and the ratio $T/|U|$ is computed by {\sc rockstar}. 
We also consider halo concentrations, but these tend to be poorly defined for subhalos. 
Therefore, to estimate the concentration parameter, we adopt the following formula which is a more robust estimate than $r_{vir}/r_s$ \citep{Springel:2008gd,Dooley:2014db} which we denote as $c_{\rm vmax}$:
\be
\frac{c^3}{\ln(1+c)-c/(1+c)}=0.216(\frac{V_{\rm max}}{H_0 r_{\rm vmax}})^2
\ee

\subsection{Selecting UFDs and their progenitors}

To trace UFD progenitors, we must first identify a population of dark matter halos at $z=0$ as UFD host candidates. Ideally this would be done by constructing a physical model that predicts the luminosities of all subhalos. However, the relevant physics for forming low-luminosity galaxies is still not completely certain \citep[e.g., ][]{Lu:2017fa}, and it is still challenging to directly simulate a Milky Way galaxy while resolving its satellites \citep{Fattahi:2016dh}.

\begin{figure}
\includegraphics[width=\linewidth]{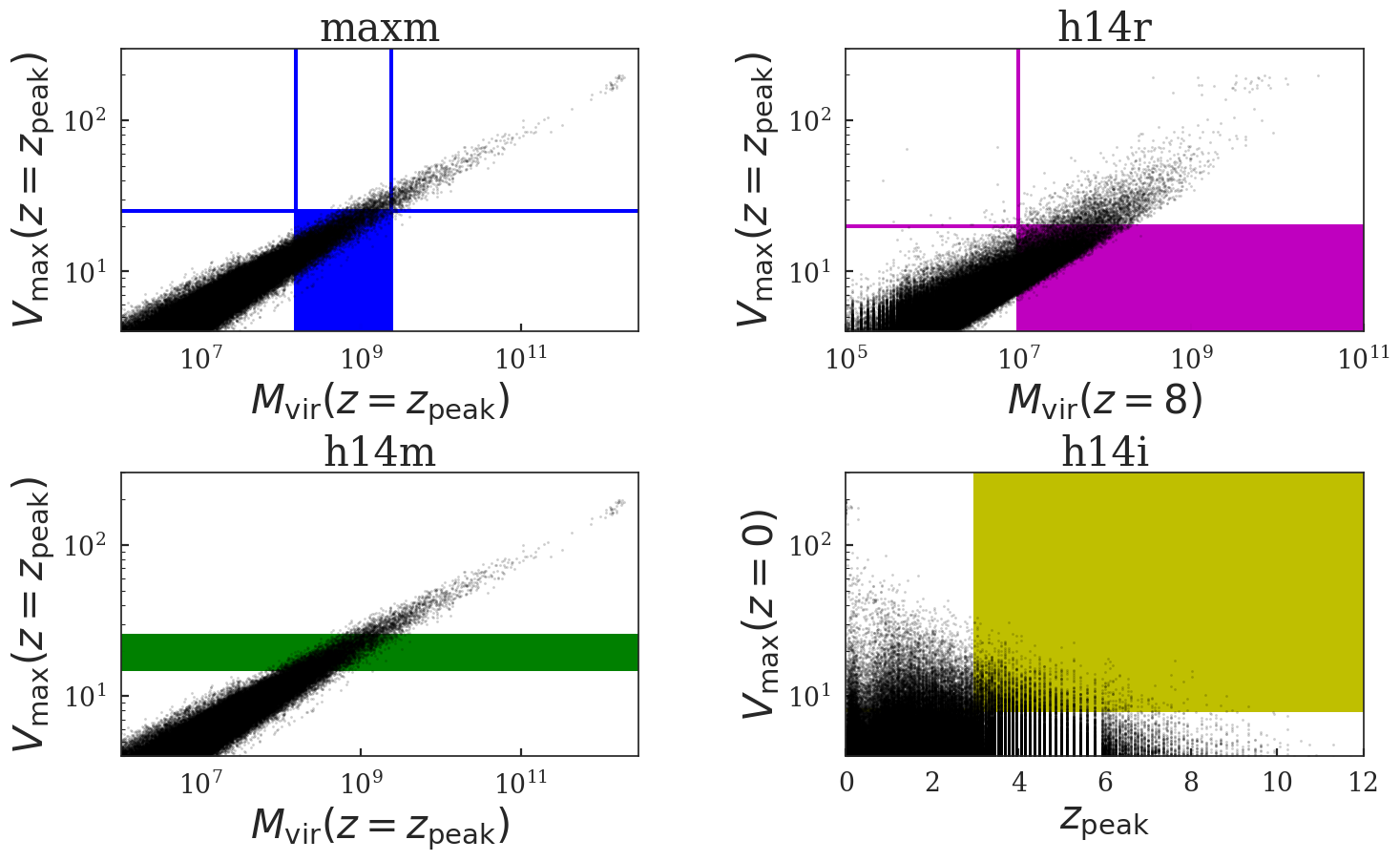}
\caption{Selection criteria for UFDs. Each panel shows all halos at $z=0$ in the main host (black points), as well as the selection cuts defined to select UFDs (colored lines and shaded region; the same colors are used for later plots).
Model h14i also has $\vpeak < 25\kms$ applied.
\label{f:ufdselection}}
\end{figure}

Various methods have been adopted to identify UDF candidates from high resolution N-Body simulations based on abundance matching (AM) techniques \citep{Moster:2013dl,Behroozi:2013fga,GK14a,GarrisonKimmel:2017cv,Brook:2014dv,Jethwa:2016uj}. Given the method of choice, the statistics of the UFD candidates such as their number density and spatial distribution can also vary \citep{GK14a,Hargis14,Dooley:2016vt}. 
Here, we choose four simple definitions based on a halo's mass and accretion history, which are shown in Figure~\ref{f:ufdselection} and described as follows: 
\begin{itemize}
\item \maxm: Surviving $z=0$ halos with stellar mass between $10^3 - 2\times10^{5} \msun$ using the abundance matching relation of \citet{GK14a}:
\begin{equation}
M_\star(M_{\rm peak}) = 3 \times 10^6 M_\odot \left(\frac{M_{\rm peak}}{10^{10} M_\odot}\right)^{1.92}
\end{equation}
The luminosity range corresponds to $10^{8.2} \msun \lesssim \Mpeak \lesssim 10^{9.4}\msun$. 
This relation is valid only at z = 0 and does not allow for a constraint at higher redshifts (See \citet{Deason:2016hh} for a discussion of why this relation does not evolve with redshift).
\item \hargism: Massive in the past : Surviving $z=0$ halos with $\vpeak > 15\kms$ \citep{Hargis14}. These halos are \emph{not} required to have formed prior to reionization.
\item \hargisr: Formed before reionization: Surviving $z=0$ halos with $\vpeak < 20\kms$ and $M_{z=8} > 10^7\msun$ \citep{Hargis14}. These are the halos that are formed before reionization and 
correspond to the fossil definition of UFDs \citep{Bovill:2011bk}.
\item \hargisi: Earliest infall: Surviving $z=0$ halos with $v_{\rm max, {z=0}} > 8\kms$ and $\zpeak > 3$ \citep{Hargis14}. This corresponds to associating UFDs with massive halos that 
lost most of their gas and dark matter content through stripping processes after they infall into the MW host at $z>3$ \citep{Kravtsov:2004he}.
\end{itemize}

In all UFD selections methods, we only consider objects with $\vpeak \leq 25\kms$.
The value of $25 \kms$ is consistent with the criterion of circular velocity of the halos being less than $v_c<20\kms$ \citep{Bovill:2009hg,Ricotti:2005dj}
to define pre-reionization fossils based on the combined effect of feedback and cooling in such low mass halos at redshifts of reionization.
We also note that all halos that we have defined as UFD candidates are true fossils in that they never reached a mass scale in their
entire history beyond the filtering mass corresponding to $v_{filt}=20-30\kms$ to be able to accrete gas from the IGM and therefore reignite their star formation and be considered as a polluted system \citep{Gnedin:2006gl,Bovill:2011bk}.

The main branch of a $z=0$ halo is defined by tracing the most massive progenitor at each snapshot. For each MW simulation, we trace the main branch of all halos at $z=0$ back to a higher redshift $z$. This halo at redshift $z$ is called the \emph{main progenitor} of the $z=0$ halo.
We only consider objects with mass $> 10^{6.5}\msun$ at all redshifts.
We ignore halos that do not end up within the virial radius of the $z=0$ host for two reasons. First, the Caterpillar halos are only guaranteed to be uncontaminated by low-mass-resolution particles within the virial radius. Second, nearly all galaxies in the UFD mass range are observed within the Milky Way's virial radius. 
Some UFDs are recently discovered outside the local group (LG) \citep{Lee:2017hu} which shows these systems are ubiquitous though challenging to detect because of their low surface brightness \citep[see also ][]{Wheeler15, Dooley16}.

\begin{figure*}
\includegraphics[width=\linewidth]{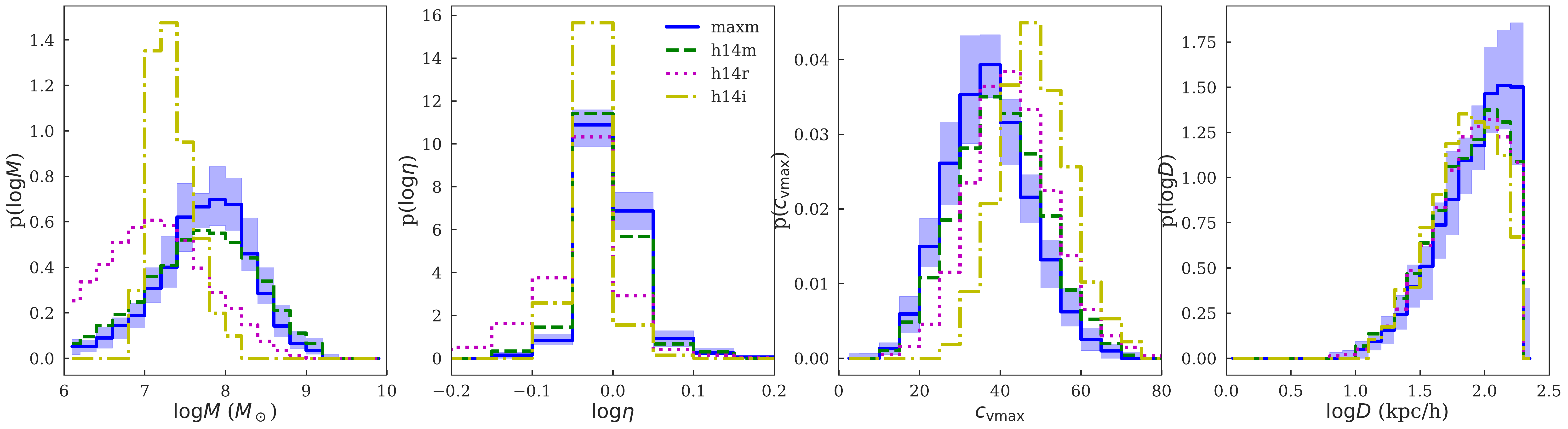}
\caption{The distribution of the UFD host halos that are selected at $z=0$ based on the above four AM techniques in mass, virial ratio, concentration parameter, and distance from the main progenitor. 
The histograms are normalized to better illustrate the differences between the models. The most important differentiator between the models is the mass distribution of the UFD host halo candidates.
\label{f:z0}}
\end{figure*}

The distribution of the UFD candidates that are selected at $z=0$ based on the above four AM techniques for different dynamical parameters is shown in Figure \ref{f:z0}.
The histograms are normalized to better reflect the distributions.  The largest difference between the models is reflected in the mass distribution of the UFD halo candidates at $z=0$.
The distributions in concentration and radial distance from the center of the halo is about the same for all four models. In all cases the results of the \hargisi~model is more distinct than those of the other models
as halos in this model have been processed more efficiently through tidal stripping. 

\section{Results}

\subsection{Distribution of halos in various dynamical parameters}

The first question we address is: For surviving UFDs today, what is their distribution in various parameters that describe the halo's dynamical status at high redshifts? We study mass, concentration parameter, virial ratio and the distance from the main progenitor. By surviving, we mean the halo finder is able to identify the halo at z=0.

\begin{figure*}
\includegraphics[width=\linewidth]{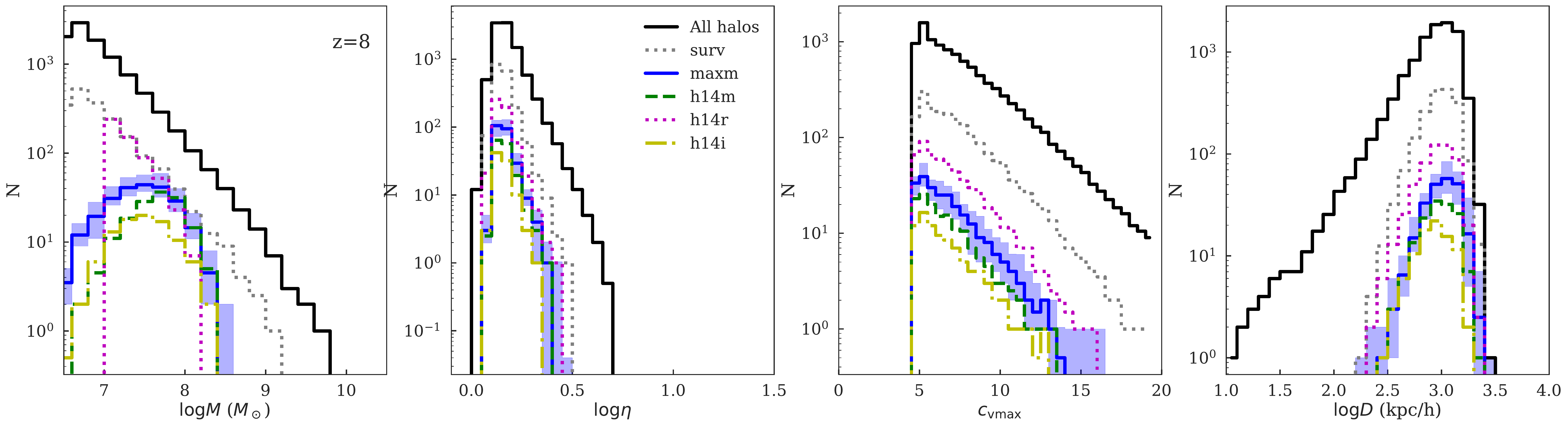}
\includegraphics[width=\linewidth]{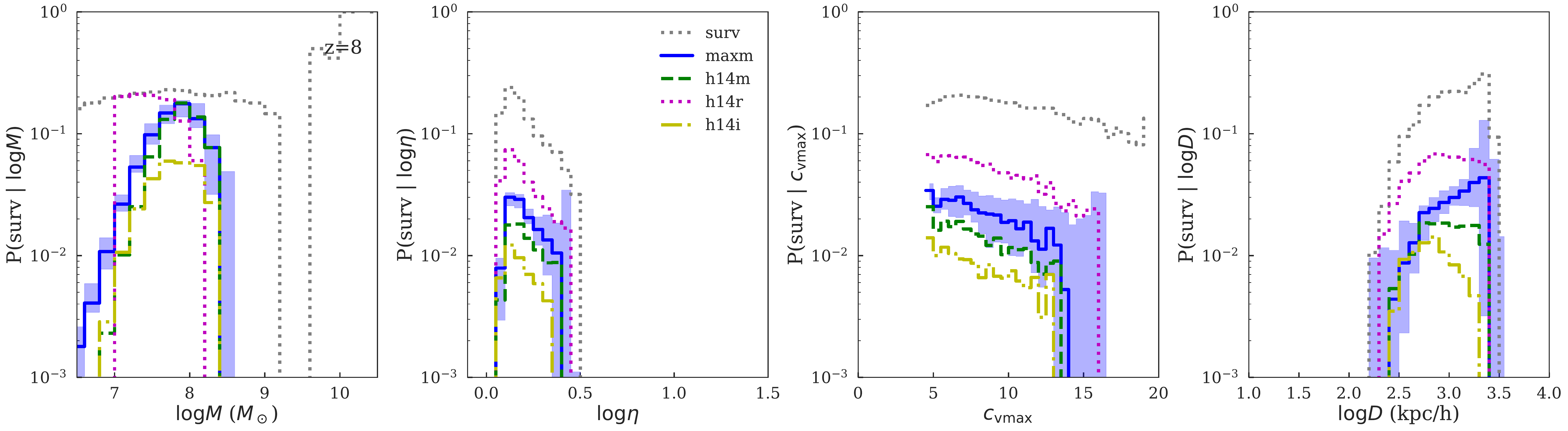}
\caption{Top: Histogram of halo masses at $z=8$ in all 32 Caterpillar simulations. Solid black line shows the histogram for all the halos at $z=8$. Dashed black line shows those halos that survive intact till $z=0$. 
Colored lines indicate objects at $z=8$ that become identified as UFDs according to four different definitions (see text).
Bottom: Fraction of halos at $z=8$ that would survive (dashed line) or be considered as UFDs (colored lines) by $z=0$. These lines are the results of dividing the colored and dashed lines in the top panel by the solid black line. 
The shaded regions correspond to 68\% MW-to-MW halo scatter. 
\label{fig:1}}
\end{figure*}

We first show the overall distributions of the halos at $z=8$ for these parameters and compare them to the progenitors of the UFD candidates. 
The top row of Figure \ref {fig:1} shows the number of all MW and MW's satellites progenitor halos at $z=8$ for all the 32 {\it Caterpillar} halos as a solid black histogram. The dashed lines show all those that survive until $z=0$. The other colored histograms show, for each parameter, the distribution of those that survive as a satellite UFD candidate (based on four different methods to identify them) at $z=0$.
The bottom row of Figure \ref{fig:1} shows the ratio of the colored and dashed histograms in the top panel to the solid black one. 
Therefore, the bottom panel of Figure \ref{fig:1} shows the conditional probability that a halo at $z=8$ survives \emph{intact} as a UFD (colored lines) or ends up as any surviving subhalo (black dashed line) given its mass, virial ratio,  concentration parameter, and distance from the main progenitor. 
In all panels, the shaded regions represent the 68\% scatter between the 32 MW host halos in {\it Caterpillar} (this is effectively the same as dropping the highest and lowest value in each bin).

When showing the distribution in mass, we set a lower limit to $\log (M/\msun)\approx6.5$ because that corresponds to a halo with at least 100 DM particles. 
We note that by performing hydrodynamical simulations coupled with radiative cooling on individual halos, \citet{BlandHawthorn:2015ke} arrive at a physical lower mass limit for a UFD 
progenitor of $\approx10^7\msun$. This mass limit is for an isolated halo to survive a SN explosion in the presence of radiative cooling. The survival mass limit could be affected if modeled in a cosmological setting due to constant accretion of the cold gas from cosmic web.

Figure \ref {fig:1} shows that halos have the highest probability to survive as UFD defined in the \hargisr~ model compared to other three models to identify UFD candidates at $z=0$. 
The reason being simply that there are more halos at $z=0$ that satisfy the \hargisr~model definition of a UFD host candidate.
The lowest probability is associated with the \hargisi~model. This is also shown in \citet{Hargis14}. The reason behind this is that in the \hargisi~model, we are focusing on those halos that have entered their host halo at $z>3$ and therefore have had a longer time of experiencing the tidal forces in the host leading to their destruction. However, we note that 
for the \hargisi~case the criterion is a limit on the maximum circular velocity at $z=0$ as opposed to the other three models where the cut on $\vmax$ is imposed at $\zpeak$. Though we have imposed a limit for a halo to be identified with at least 100 DM particles at high redshifts, the limit at $z=0$ is only 10 particles which is the default of the halo finder algorithm. Imposing a 100 DM particle at $z=0$ similar to high redshift, would reduce the number of candidates at $z=0$ for the other three models and bring the four models into better agreement with each other.

The population of all the surviving halos (regardless of whether they end up as a UFD candidate or not) and those that survive as UFDs have similar distribution in distance from the host,
concentration, and virial ratio, while they are noticeably different in mass. The survival probability for all halos has a flat distribution in mass when we only care about whether they survive as a MW satellite or not.
The chances of survival as a UFD candidate peaks around $\log (M/\msun)\approx7.5-8$ at $z=8$. Therefore, mass is the desirable parameter if we are about to select
halos at $z=8$ to have a higher chance to become UFD candidates as opposed to merging into or surviving as a classical dSph galaxies.
Note that we selected UFDs primarily by their halo mass, so essentially there is a high correlation between $M(z=8)$ and $\Mpeak$). Indeed around $\log (M/\msun)\approx7.5-8.5$ almost all the survived halos are 
those that survive as a UFD candidate as the difference between the dashed line and the colored lines is shrunk to zero around that mass range (see bottom panel of Figure \ref {fig:1}, the distribution in mass).
The means by which the knowledge of other parameters describing a halo at high redshifts would inform us about the survival probability is explored in Section \ref{ss:other_parameters}.

It should be noted that between different AM techniques to identify the UFDs at $z=0$, the \maxm~method, which is an observationally oriented technique, leads to a distribution that is in better agreement with 
the predictions of \hargism~(massive in the past) model. In other words, comparing colored lines in the bottom panel of Figure \ref{fig:1}, the \hargism~ line is closer to \maxm~ line in all the parameters studied. 
This could also be read from Figure \ref{f:ufdselection} where one sees that the selection criteria for the \maxm~and \hargism~bracket almost the same parameter space.

\subsection{Fate of a halo given its mass}

We now turn the question around and ask what is the fate of halos at high redshifts given their mass and specifically what is the probability of surviving as a UFD candidate? 
We trace all the halos in the simulation box at $z=8$ (shown as the dashed black histogram in the top panel of Figure \ref{fig:1}) down to $z=0$ and study
their fate at $z=0$ given their mass at $z=8$. 
We define five different $z=0$ categories, tracking halos that:
\begin{itemize}
  \item{Merge into their hosts.}
   \item{Merge into classical dSph galaxies, defined as halos with $M_\star > 2 \times 10^5 \msun$, or $M_{\rm peak} > 2.5 \times 10^{9} \msun$, 
   according to the \citet{GK14a} abundance matching model.}
\item{ Become the main progenitors of classical dSph galaxies.} 
\item{ Merge into UFD candidates.}
\item{ Become the main progenitors of UFD candidates.}  
\end{itemize}
Halos merging into objects with mass range less than UFDs or equal to UFDs but  not satisfying the UFD conditions are not studied.

\begin{figure*}
\centering
\includegraphics[width=\linewidth]{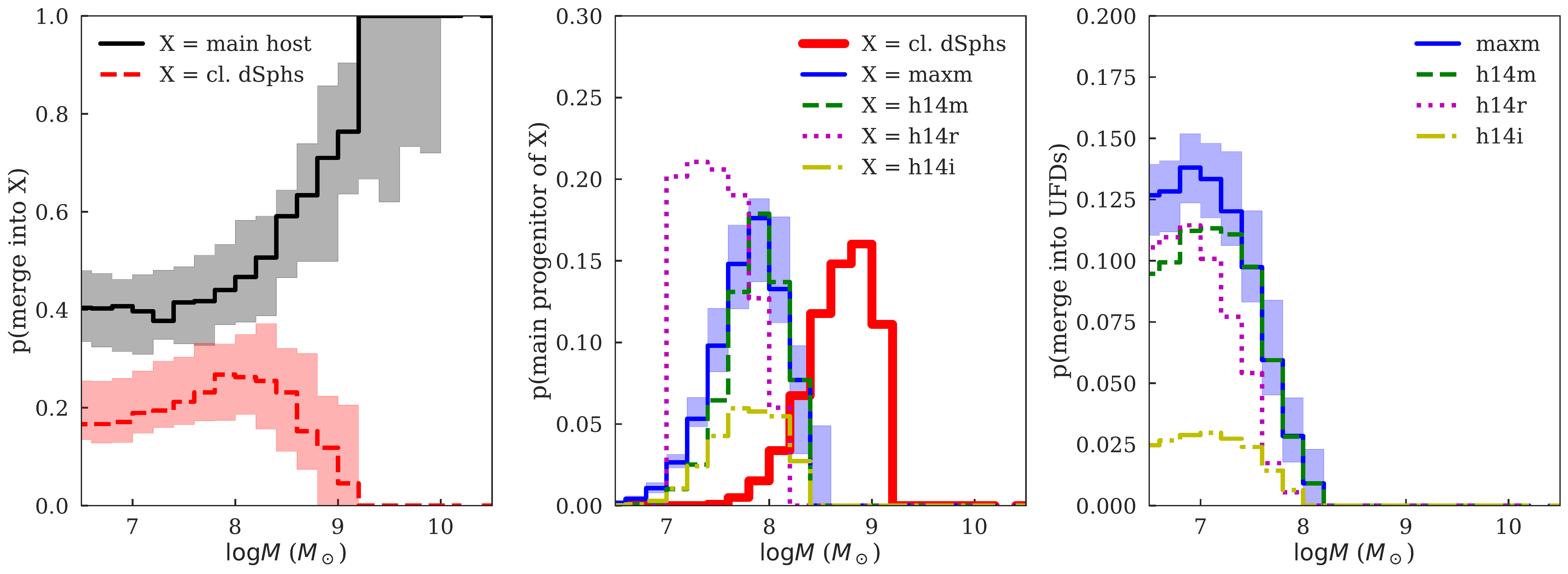}
\caption{The fate of halos at $z=8$ as a function of their mass at $z=8$.
Panel (a): At a given mass bin the fraction of $z=8$ halos that merge into the main host and a classical dSph galaxy \citep[defined by $M_\star > 2 \times 10^5 \msun$ using the abundance matching model of][]{GK14a}. Panel (b): shows the probability of becoming the main progenitor of a classical dSph galaxy (thick red line) and a surviving UFD at $z=0$. The solid-blue,  dashed-green, dotted-pink and dash-dotted yellow lines correspond to \maxm, \hargism, \hargisr~and \hargisi~models respectively.
Panel (c):  The probability that a $z=8$ halo has merged into a surviving UFD but is \emph{not} the main progenitor of the UFD. For example, ${\approx}40\%$ of $10^{7.5}\msun$ halos at $z=8$ have merged into the main host, at the same mass bin ${\approx}20\%$ merge into other classical dSph, ${\approx}5-15\%$ survive as UFDs, and less than 10\% merge into a UFD candidate at $z=0$. In all panels, the lines indicate the median in each mass, shaded bars indicate 68\% halo-to-halo scatter.
\label{fig:2}}
\end{figure*}

Figure \ref{fig:2} shows the fate of halos at $z=8$ as a function of their mass for those five categories.
In Panel (a) we show the fraction of halos that would merge into the main host and classical dSph galaxies as a function of their mass. The larger probability to merge into the main host for more massive halos is due to the dynamical friction \citep{Chandrasekhar:1943vr}, which is more effective for more massive satellites. Nearly 40\% of halos with $M_h\sim10^{7}\msun$ merge into the main host, and the probability rises to unity for halos with mass $M_h>10^{9}\msun$. 
Panel (b) shows the probability of becoming the main progenitor of a classical dSph galaxy (thick red line) and a UFD candidate based on our four different AM techniques to identify them at $z=0$. A similar result has been shown in \citet{Bovill:2011bk} where fossils formed in biased regions will merge into non-fossil halos with circular velocities $v_c>20-30\kms$. 
\citet{Bovill:2011bk} found that while there was some dependence of survivability of a fossil on mass, the strongest dependence was on the luminosity of the fossil, likely because the most luminous fossils were forming in slightly over-dense regions and therefore more likely to merge.
This result is also consistent with \citet{Gnedin:2006gl} where they find 5-15\% of dwarf satellite galaxies at $z\approx 8$ survive until the present without significant evolution since then.
Halos with mass between $\log (M/{\msun}) \approx 7-8$ have about 10-20\% chance of surviving as a UFD based on three of the four different selection cuts (\hargism,\hargisr, and \maxm).
The yellow line (\hargisi~model) that corresponds to the earliest infall model \citep{Kravtsov:2004he} shows different results compared to the other three AM methods. \hargisi model only include systems that are accreted into the main host at $z>3$ and therefore they experience tidal stripping for a rather long time compared with the other three models.
Panel (c) shows the probability that a halo merges into a UFD candidate at $z=0$, which naturally peaks at lower masses compared to what is shown in panel (b). 
The shaded regions are 68 percentile computed from all the 32 MW halos modeled in {\it Caterpillar} simulation. 

\begin{figure*}
\begin{center}
\includegraphics[width=\linewidth]{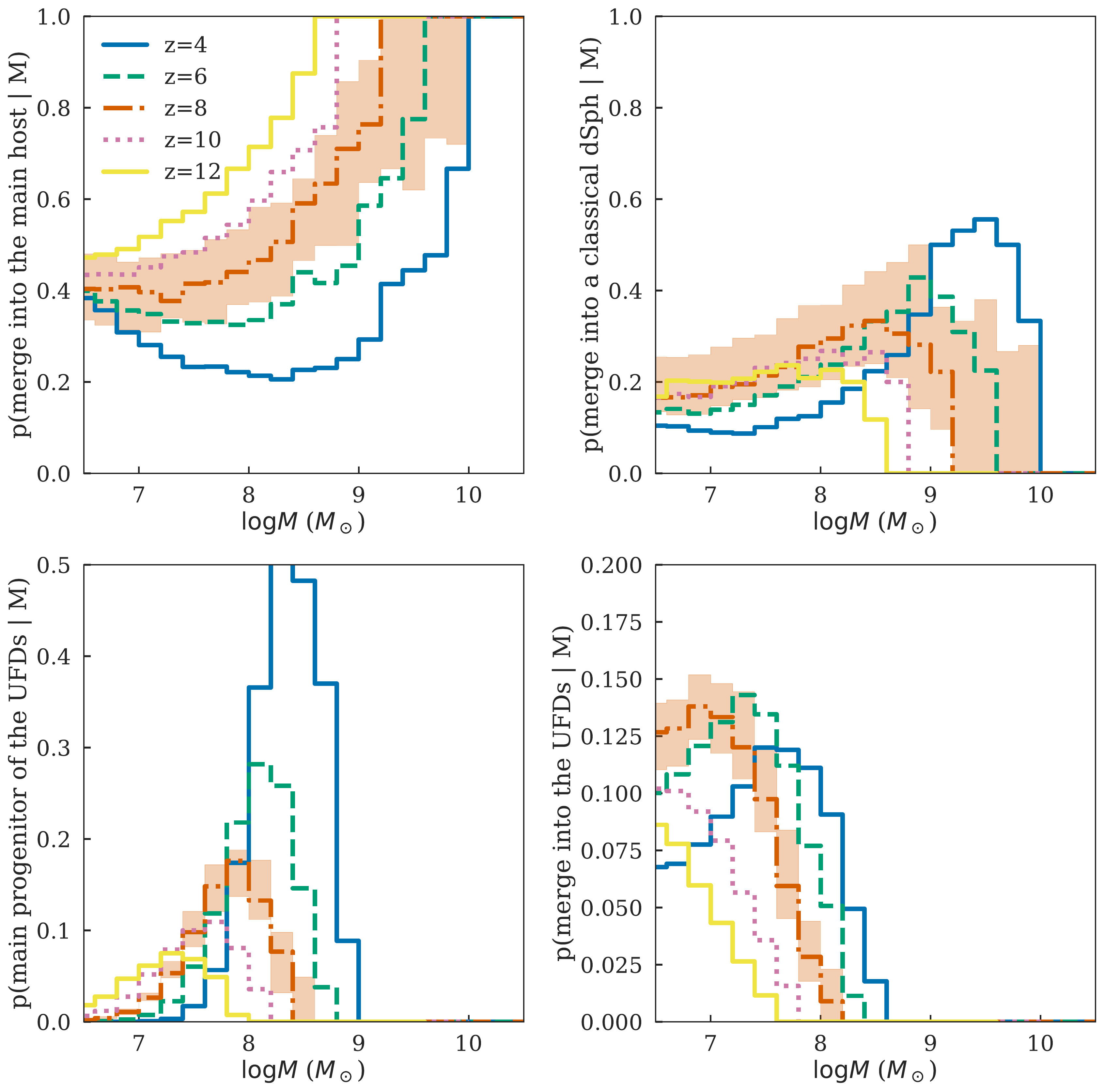}
\end{center}
\caption{Same as Figure~\ref{fig:2} but shown for different redshifts. Here we only use the ``\maxm'' model to identify the UFD candidates. Panel (a): At a given mass, higher redshift halos have a higher probability to merge into the main host. For example a $10^{8}\msun$ halo at $z=12$ has a 60\% chance of merging into the main host halo while the same mass halo at $z=4$ has 20\% chance to merge into the main host. The reason being the mass accretion of halos such that a $10^{8}\msun$ halo at $z=12$ would reach masses around  $10^{10}\msun$ by $z=4$. 
The general trend of more massive halos being more prone to merging is because of dynamical friction. See text for explaining the reason behind the upturn at low masses towards low redshifts. Panel (b): The probability of merging into a classical dSph galaxy becomes larger for more massive halos and at a given mass bin above $>10^{8.5}\msun$ halos at lower redshifts have a higher probability to merge into a classical dSph galaxies while the situation reverses going to lower mass bins. Panel (c): The probability of being the main progenitor of a surviving UFD depends on mass and redshift of the halo. The peak of the probability shifts from lower masses at higher redshifts to higher masses at lower redshifts. Panel (d): The same trend is observed for halos merging into a surviving UFD.
\label{fig:3}}
\end{figure*}

We apply the same procedure at other redshifts (at $z=4,6,8,10,12$).
In panel (a) of Figure \ref{fig:3} we show the probability of merging into the main host as a function of redshift for a given mass bin.
For a halo at given mass, being at higher redshifts increases its probability of merging into the main host. 
This is intuitive as a halo at higher redshift will grow in mass from $z\approx 12$ to $z\approx 4$ and therefore a smaller halo at $z\approx 12$ (for example with mass $\approx 10^8 \msun$) would have a similar fate as a more massive halo ($10^9 \msun$) at $z\approx 8$. Although formation of halos in less biased regions can lead to higher survival probability; we show in Figure \ref{fig:6} that this is weakly dependent on mass.
The upturn at low masses which is pronounced in the $z=4$ curve, is a numerical artifact. The halo finder has a difficult time recovering sub halos with large mass ratios between the sub halo and the host, specifically at small distances from the host.
This is illustrated in \citet{Behroozi:2013cn} (See their Figure 10) for a massive host halo in which satellites with small satellite mass ratio are not recovered at small distances from the host. 
Low mass halos at high redshifts (for example at $z=12$ ) merge into a smaller host and the halo finder does not face as much difficulty in correctly tracing them in time.

However, at redshifts $z\approx4$, the same low
mass halos tend to merge into more massive hosts and therefore having smaller satellite-to-host mass ratios. Due to this fact, the merging of these small halos is artificially boosted to higher values contrary to what one would expect from the dynamical friction arguments. 
Panel (b) shows the redshift evolution of the halos merging into a classical dSph galaxy, which shifts towards larger masses at lower redshifts.  
Panel (c) shows that the probability of a halo to survive as a UFD shifts towards larger masses at lower redshifts. We have selected ``maxm'' model as our fiducial case. For example, a halo with mass $\approx 10^{8.5} \msun$ has zero 
chance to survive as a UFD candidate at $z=12$ while halos of the same mass at $z=4$ have 40\% chance of surviving as a UFD candidate.
Panel (d) shows the probability to merge into a UFD candidate as a function of redshift which drops at higher masses at all redshifts. The probability drops to zero
at the mass scale where being the main progenitor of the UFD peaks. For example at $z=4$, halos with mass $\approx10^{8.5}\msun $ have the highest chance of being the main progenitor of the UFD candidates and
the probability distribution of the other progenitors drops to zero for masses above $\approx10^{8.5}\msun$. Meanwhile, halos as small as $\approx10^{7}\msun$ have no chance of being the main progenitor of a UFD at $z=4$, while they can be merge into the main progenitor of a UFD with probability of around 10\%. 

\subsection{Informants other than mass}\label{ss:other_parameters}

\begin{figure*}
\begin{center}
\includegraphics[width=0.8\linewidth]{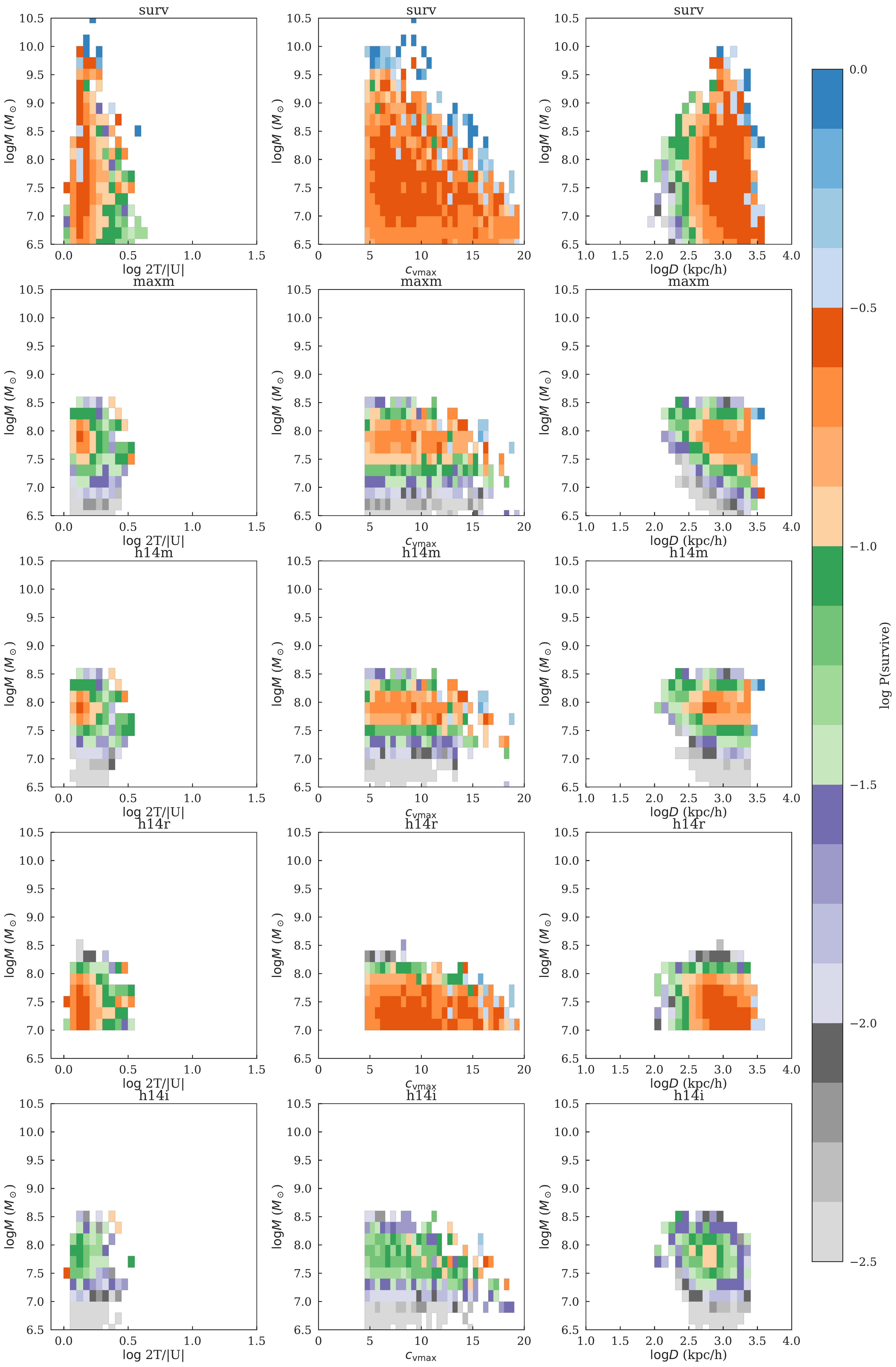}
\end{center}
\caption{The survival probability of all the halos as UFDs in the 2D space of mass \emph{and} virial ratio (first column), concentration (second column) and distance from the main progenitor (third column) at $z=8$. 
The color coding is showing the probability of survival. Each row corresponds to a different AM technique to select satellite UFD candidates at $z=0$. At a given mass bin, selecting halos with lower
virial ratio and with larger distance from the main progenitor boosts their chance of survival as UFDs by a factor of $\approx3$. The trend with concentration parameter remains inconclusive. 
\label{fig:6}}
\end{figure*}

In this section we study whether knowledge of dynamical parameters of the halos other than their mass can be useful in predicting their survivability as UFD hosts today.
Figure \ref{fig:6} shows the survivability in the 2D space for all the halos (in the top row) and for UFDs given the AM technique by which they are identified. 
Each row shows a different method to identify UFD candidates at $z=0$ where the top row indicates the survival chance for all halos.
We see a trend with lower virial ratio and larger distance from the main progenitor. Whether the concentration parameter affects
the halo's survival remains inconclusive. 
Note that the AM technique adopted in the \hargisi~ predicts less of a trend with virial ratio (lower probability at the same ($M, \log \eta$) ) bin
compared to the three other AM techniques. As we will discuss in Section 4, lower virial ratio is indicative of higher formation redshift or no recent merger activity and in the \hargisi~model it is not the oldest that survives preferentially.
The probability of survival of halos in the \hargisi~ model is on average about 1/3 of the other three AM techniques.
At a given mass bin, one can select halos with low virial ratios or large distances from the main host to potentially boost the survival chance by a factor $\approx3$.

\begin{figure*}
\begin{center}
\includegraphics[width=\linewidth]{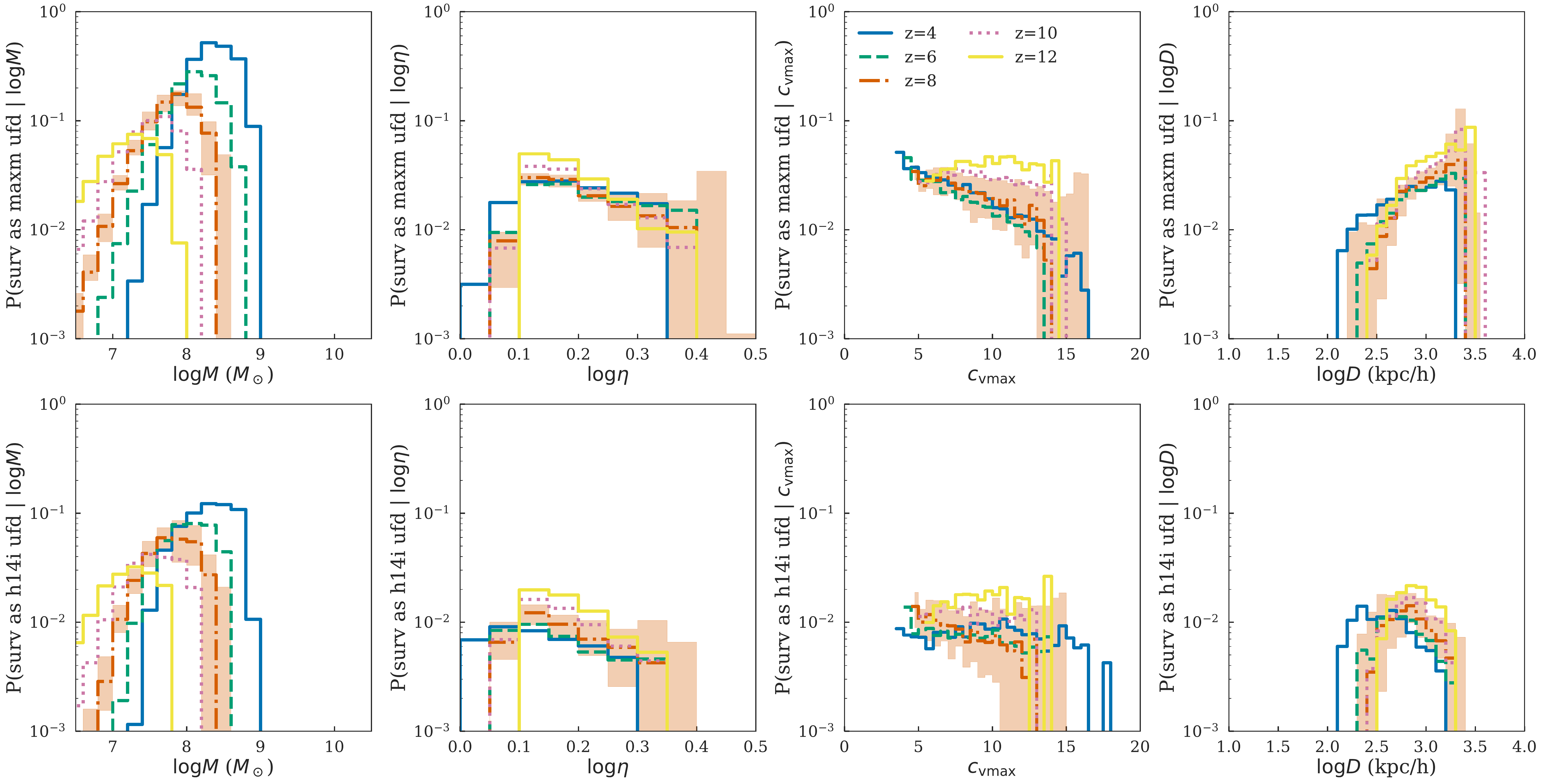}
\end{center}
\caption{The trend of survival probability as a UFD at a given redshift for different dynamical parameters. Top row: \maxm~method is adopted to define a UFD at $z=0$. 
The data suggests that the halos with higher concentration preferably survive at higher redshifts and it switches
to those with lower concentration as we go to lower redshifts. The trend with mass is discussed before and the trend with other parameters remain less dependent on the redshift of consideration. 
Bottom row: \hargisi~method is adopted to define a UFD at $z=0$. The same trends as in the top row are observed, however, the concentration is flatter at all redshifts. 
In other words, the \hargisi~method shows no preference in selecting halos if it is only based on the concentration parameter. 
\label{fig:7}}
\end{figure*}

Figure \ref{fig:7} shows the trends with redshift to see whether a parameter becomes a stronger informant at higher or lower redshifts compared to the $z=8$ results presented above. 
We show the results for two different methods of selecting UFDs at $z=0$ (\maxm~in the top row and \hargisi~in the bottom row). 
The left panel of Figure \ref{fig:7} shows the probability to survive as a UFD as a function of mass at different redshifts. At lower redshifts, the preferred mass
range with the highest chance of survival increases to larger masses. For example, halos in the mass range $\log (M/\msun) =7-7.5 $ at $z=12$ have the highest chance of survival 
while halos in the mass range $\log (M/\msun) =8-8.5 $ at $z=4$ have the highest probability to survive as a UFD identified by the \maxm~method.
The data suggests that for the \maxm~method of defining UFDs, halos with higher concentration preferably survive at higher redshifts and it switches
to those with lower concentration going to lower redshifts. There is no trend with concentration parameter if for example the \hargisi~method is applied to identify the UFD candidates at $z=0$. 
There appears to be no redshift trend with virial ratio. Moreover, being at larger distances seems to increase the chance of survival more effectively at higher redshifts. 

\section{Summary and Discussion}

The survival of subhalos has been extensively studied in the literature, and the results all suggests that 
 the satellite-to-halo mass ratio  and the orbit of the satellite are the key factores \citep{Gnedin:2006gl,Sales:2007kw,BoylanKolchin:2008ex,Penarrubia:2008kp,Klimentowski:2010fu,Bovill:2011bk,GarrisonKimmel:2017dy}. Yet, although the physical mechanism that determines the survival of the halos is well explored, the probability of surviving until $z=0$ given a halo's mass and other characteristics at higher redshifts is poorly studied.   \citet{Gnedin:2006gl} show that  5\%-15\% of subhalos in the mass range corresponding to dSph satellites at $z\approx 8$ survive until the present in a MW-like environment. 

In this work, we extend these studies to the realm of UFDs. Our results are based on analysis of merger trees constructed from the {\it Caterpillar} simulation suite \citep{Griffen16a, Griffen16b} which, consists of 32 MW type halos simulated with a mass resolution of $m_p\approx10^{4}\msun,$ suitable to study UFD progenitors.  We identified UFD candidates at $z=0$  based on four selection criteria discussed in \citet{Hargis14,GK14a}, and trace all the halos in the zoom-in simulation box that will end up within the virial radius of a given MW type halo from high redshifts down to $z=0.$ 

We find that a large fraction of the progenitor halos merge into dSph like subhalos or into the main host  due to dynamical friction, but about 10\% of the halos at $z=8$ with $\log M/\msun=7-8$  survive to the present day as UFD host candidates.  At a given halo mass, the probability of survival decreases towards higher redshifts, because those halos represent more massive halos at lower redshifts, and are therefore they are more likely to merge into large subhalos or into the main host. We  also consider the effect of three other parameters on UFD survival probability: concentration, virial ratio and distance from the main progenitor. Our results suggest that choosing halos with lower values of the virial ratio, which is indicative of higher formation redshift and a lack of recent merger activity \citep{Power:2012js,Cui:2017gd} or selecting halos at large comoving distances from the main progenitor ($D>300 {\rm kpc/h}$), which is indicative of longer infall time, can boost the survival probability by a factor of $\approx3$. The behavior with concentration parameter remains inconclusive, but the results suggest slightly higher chances of survival for halos with higher concentration if selected at higher redshifts, and lower concentration if selected at lower redshifts. 

Three main uncertainties remain that could affect our N-body results.  The first is the presence of baryons, which have the potential to have significant effects on the underlying dark matter distribution. This has been explored in other studies \citep{Geen:2013kf,Despali:2017iw,Sawala:2017jp,GarrisonKimmel:2017dy,Errani:2017jx}. For example, \citet{Sawala:2017jp} explored the impact of including baryons on the statistics and spatial  distribution of the satellites of a MW type halo in the mass range of $10^{6.5}-10^{8.5}\msun$. They find the number of the satellites would be reduced by 25-50\% independent of the satellite mass but that this effect would be increased toward the halo center.
Similar results are presented in \citet{GarrisonKimmel:2017dy} where by analyzing the Latte simulation suite \citep{Wetzel:2016iy}, 
demonstrate the presence of the galactic disk as the main reason why baryons impact the statistics of the small satellites compared to DM-only simulations. 

A second uncertainty is the limit of the minimum subhalo mass that is able to host a MW satellite galaxy  \citep{Dooley:2016vt,Jethwa:2016uj}. Halos below the atomic cooling limit, which corresponds to a mass $\approx10^8 \msun$ during reionization, cool via $\rm H_2$ and fine structure line cooling and their stellar mass content is sensitive to the effect of feedback from their first generation of stars. The Far Ultraviolet (FUV) radiation from the first generation of stars can destroy the $\rm H_2$, which is the primary coolant \citep[negative feedback;][]{Machacek:2001fq,OShea:2008kt}, or help the cooling efficiency through the formation of $\rm H^{-}$ \citep[positive feedback; ][]{Haiman:1996ii}. Such considerations can theoretically constrain the lower limit on the host halo mass of UFDs that could survive until present day. However, below $M_h<10^8\msun$, star formation becomes a stochastic process that increases the scatter in the $M_\star-M_h$ relationship and leads to different behavior  for halos below the atomic cooling limit \citep{Chen:2014co,GarrisonKimmel:2017cv}. For example, studying individual halos through a set of hydrodynamical simulations, \citet{BlandHawthorn:2015ke} showed that the lowest possible mass of a UFD progenitor is $\approx10^7\msun$, as halos below this mass cannot survive the energy input of a single SN explosion, despite the presence of radiative cooling. On the other hand, by performing zoom-in hydrodynamical simulations on galaxies and UFDs, \citet{Wheeler:2015fm} concluded that  $M_h\approx5\times10^9\msun$ is minimum halo mass threshold, while \citet{Jeon:2017wo} performed cosmological simulations of individual LG galaxies within the LG to show that $M_h\approx10^9\msun$ form 90\% of their stars prior to the reionization. 

Finally, better constraints on UFD progenitor halo could be determined if the $z=0$ properties of UDF halos were better known.
Though UFDs are likely ubiquitous in the universe \citep[e.g.][]{Lee:2017hu},  their low luminosities make them currently observable only as Milky Way satellites. Even at these distances, poorly estimated incompleteness limits our determination of their luminosity function and radial distribution \citep[e.g.][]{Koposov:2008ja,Newton:2017tn}. Their color magnitude diagrams and chemical abundance distributions of UFDs clearly point to systems that formed all their stars within the first two Gyr of the universe \citep{Vargas:2013ei,Brown:2014jn,Weisz:2014cp,Frebel:2014gi}, but the masses, concentrations, and virial parameters of dark matter halos they contain are poorly constrained due to the relatively small number of stars currently accessible with spectroscopy. Thus, direct constraints on UFD dynamical masses are restricted to the stellar half light radius \citep{Wolf:2010df}, and inferring the total halo mass of a UFD requires extrapolating an assumed density profile beyond the stellar observations \citep[e.g.][]{Strigari:2008in}.

Despite these issues, our results can be used as a rough guide to those performing zoom-in simulations on UFD candidates \citep[e.g.][]{Safarzadeh:2017dq,Jeon:2017wo}, as they allow for analyses of a limited sample of halos at high redshifts without the need to trace the simulations down to $z=0.$  Given the survival statistics presented in this paper, one can carefully select high-redshift halos, based on their mass and other characteristics, and study them in detail knowing the probability of them to survive as UFD candidates at $z=0$. \citet{BoylanKolchin:2016ex} have shown that the LG progenitors at redshifts $z>8$ have a linear comoving size of $\approx 8$ Mpc/h and are large enough to be cosmologically representative in terms of  matter density and counts of dark matter halos with $M_{\rm vir}(z = 7)< 2\times10^9\msun$. Therefore, if one simulates a cosmological volume of similar size down to the reionization redshift, that volume could end up hosting a LG at $z=0$. 

Adopting the survival statistics of the halos that we present above would imply that, out of $N$ halos selected from such simulations at a given mass bin and redshift with survival probability of $P_s(M,z)$, the chances that at least one of them become a UFD host candidate at $z=0$ is $P=1-(1-P_s(M,z))^N$.  Therefore if we select 20(10) halos with survival chance of 10(20)\%, there is $\approx$90\% chance that at least one of them will end up as a UFD host candidate at $z=0$. Considering other characteristics that boost the survival as UFD candidates by a factor of $\approx3$ would lead to only a handful of halos needed to be simulated at high redshifts to confidently generate at least one $z=0$ UFD host candidate. 
Future work on this topic, including baryonic effects and studying other halo properties, may be able to improve these probabilities further, boosting the reliability of near field cosmological studies by better selecting ultra-faint dwarf candidate progenitors in high redshifts cosmological simulations.

\section{Acknowledgements}
We are thankful to Mia Bovill, Brendan Griffen, John Wise and Daniel Weisz for useful discussions. 
MTS is supported by the National Science Foundation under grant AST14-07835 and by NASA under theory grant NNX15AK82G. 
APJ is supported by NASA through Hubble Fellowship grant HST-HF2-51393.001 awarded by the Space Telescope Science Institute, which is operated by the Association of Universities for Research in Astronomy, Inc., for NASA, under contract NAS5-26555.
This research is supported in part by the National Science Foundation (NSF; USA) under grant No. PHY-1430152 (JINA Center for the Evolution of the Elements).  B.W.O. was supported by the National Aeronautics and Space Administration (NASA) through grant NNX15AP39G and Hubble Theory Grant HST-AR-13261.01-A, and by the NSF through grant AST-1514700.  Computational support for the Caterpillar simulations were provided by XSEDE through the grants (TG-AST120022, TG-AST110038). We further acknowledge support of the computer cluster of the MIT Astrophysics Division, which was built with support from the Kavli Investment Fund administered by the MIT Kavli Institute for Astrophysics and Space Research.

\bibliographystyle{mnras}
\bibliography{biblio,the_entire_lib}

\end{document}